\documentclass[sn-mathphys]{sn-jnl}
\jyear{2022}%

\usepackage{siunitx}

\begin{document}

\title[VGOS VLBI Intensives between \textsc{macgo12m} and \textsc{wettz13s}]{
\phantom{a}\par\vspace{-12ex}\par
VGOS VLBI Intensives between \textsc{macgo12m} and \textsc{wettz13s} for the rapid determination of UT1-UTC}

\author*[1]{\fnm{Matthias} \sur{Schartner}}\email{mschartner@ethz.ch}

\author[2]{\fnm{Leonid} \sur{Petrov}}\email{leonid.petrov-1@nasa.gov}
\equalcont{These authors contributed equally to this work.}

\author[3]{\fnm{Christian} \sur{Plötz}}\email{Christian.Ploetz@bkg.bund.de}
\equalcont{These authors contributed equally to this work.}

\author[2]{\fnm{Frank G.} \sur{Lemoine}}\email{frank.g.lemoine@nasa.gov}
\equalcont{These authors contributed equally to this work.}

\author[2]{\fnm{Eusebio} \sur{Terrazas}}\email{chevo@utexas.edu}
\equalcont{These authors contributed equally to this work.}


\author[1]{\fnm{Benedikt} \sur{Soja}}\email{soja@ethz.ch}
\equalcont{These authors contributed equally to this work.}

\affil*[1]{\orgdiv{Institute of Geodesy and Photogrammetry}, \orgname{ETH Zurich}, \orgaddress{\street{Robert-Gnehm-Weg 15}, \city{Zurich}, \postcode{8093}, \country{Switzerland}}}

\affil[2]{\orgname{NASA Goddard Space Flight Center}, \orgaddress{\street{8800 Greenbelt Road}, \city{Greenbelt}, \postcode{20771}, \state{Maryland}, \country{USA}}}

\affil[3]{\orgname{Bundesamt für Kartographie und Geodäsie}, \orgaddress{\street{Sackenrieder Str. 25}, \city{Bad Kötzting}, \postcode{93444}, \country{Germany}}}


\abstract{
In this work, we present a status update and results of the designated research and development VLBI Intensive program VGOS-INT-S, observed between \textsc{macgo12m} and \textsc{wettz13s} for the rapid determination of the Earth's phase of rotation, expressed via UT1-UTC. 
The main novelty of these sessions is the use of a special observation strategy, rapidly alternating between high- and low-elevation scans, enabling an improved determination of delays caused by the neutral atmosphere. 
Since 2021, 25 Intensive sessions have been observed successfully. 
In early 2022, VGOS-INT-S was among the most accurate Intensive programs with an average formal error $\sigma_{\text{UT1-UTC}}$ of \SI{3.1}{\micro s} and a bias w.r.t. IERS~C04 of \SI{1.1}{\micro s}. 
Later, the session performance decreased due to multiple technical difficulties. 
}


\maketitle

\section{Introduction}\label{sec:introduction}
Among the space geodetic techniques, Very Long Baseline Interferometry (VLBI) is able to provide the most accurate and unbiased estimates of the angle of the Earth's orientation with respect to the rotation axis, expressed via UT1-UTC. 
Since 1984, regular observing campaigns have been launched for monitoring Earth orientation parameters (EOP) including UT1-UTC. 
They are now coordinated by the International VLBI Service for Geodesy and Astrometry (IVS) \cite{Nothnagel2017}.
Most sessions of the IVS observing campaigns for EOP determination either run for 24~hours to determine the full set of EOP, or for 1~hour to determine solely UT1-UTC. 
The 24-hour programs run 2--3 times a week with a latency between observations and delivery of EOP estimates of about 15--20 days. 
The 1-hour programs on average run 2--3 times a day and the latency between observations and delivery of UT1-UTC estimates is 1--3~days. 
For that reason, these campaigns are called \textit{Intensives}.
Nowadays, a number of VLBI Intensive programs dedicated to the estimation of UT1-UTC run in parallel. 

VLBI observations commenced in 1967. 
Since then, the VLBI technique went through a number of upgrades. 
The most recent upgrade is called the VLBI Global Observing System (VGOS) \cite{Niell2018}. 
The changes in that upgrade, relevant to the present study, are faster slewing speeds of $12^\circ$ over azimuth and $6^\circ$ over elevation, combined with an increased data rate of currently \SI{8}{Gbps} distributed among four bands.
The fast slewing rates reduce the time when the antenna is slewing and thus not recording signals. 
The higher data rate allows for shorter observation times to reach the desired signal-to-noise ratio (SNR). 
Combined, this leads to a significantly increased number of scans per hour, up to 100, allowing for faster sampling of the atmosphere, which is considered one of the major error sources in VLBI. 

In the current geodetic catalogs, there are over 200 extragalactic radio sources that can be observed by VGOS radio telescopes and up to 100 sources can be observed in one hour. 
Thus, the number of combinations of sources that can be selected for observations is extremely large. 
We are interested in developing new techniques for the generation of optimal observing plans, the so-called schedules, which provide UT1-UTC with minimum errors. 
The theoretical basis of the development of an optimal schedule was described in \citep{r:sch_phd}.

In order to verify the optimized scheduling algorithm, we launched a research and development Intensive VLBI observing program, named VGOS-INT-S, on the \SI{8418}{km} long baseline between \textsc{macgo12m} and \textsc{wettz13s} in 2021. 
Station \textsc{macgo12m}, also known as Mg, is located in Western Texas, USA, and station \textsc{wettz13s}, known as Ws, is located in Northeast Bavaria, Germany. 
Here, we outline the design of the observing program and discuss preliminary results.

\section{Methods}\label{sec:methods}
The major error source in geodetic VLBI is mismodeling of the path delay in the neutral atmosphere. 
The a~priori atmospheric path delay can be computed either using a regression model of surface atmospheric pressure and air temperature or by direct integration of equations of wave propagation through an inhomogeneous refractivity field that can be derived from the output of numerical weather models. 
In both cases, the accuracy of the a~priori path delay is still insufficient and we have to estimate the residual path delay in the zenith direction from the VLBI data themselves.

During the analysis, atmospheric delays are commonly divided into a hydrostatic and a non-hydrostatic (wet) part. 
While the hydrostatic part can be modeled with sufficient accuracy, the wet part has to be estimated due to its higher variability. 
Historically, one zenith wet path delay (ZWD) per one-hour observing session was estimated. 
With the fast slewing VGOS antennas, we can develop a scheduling strategy that would allow us to estimate atmospheric path delay with segments as short as 5~minutes.
To enable a more frequent estimation of ZWD, special emphasis has to be laid on providing observations at different elevation angles within the estimation interval.

\begin{figure}
    \centering
    \includegraphics[width=\textwidth]{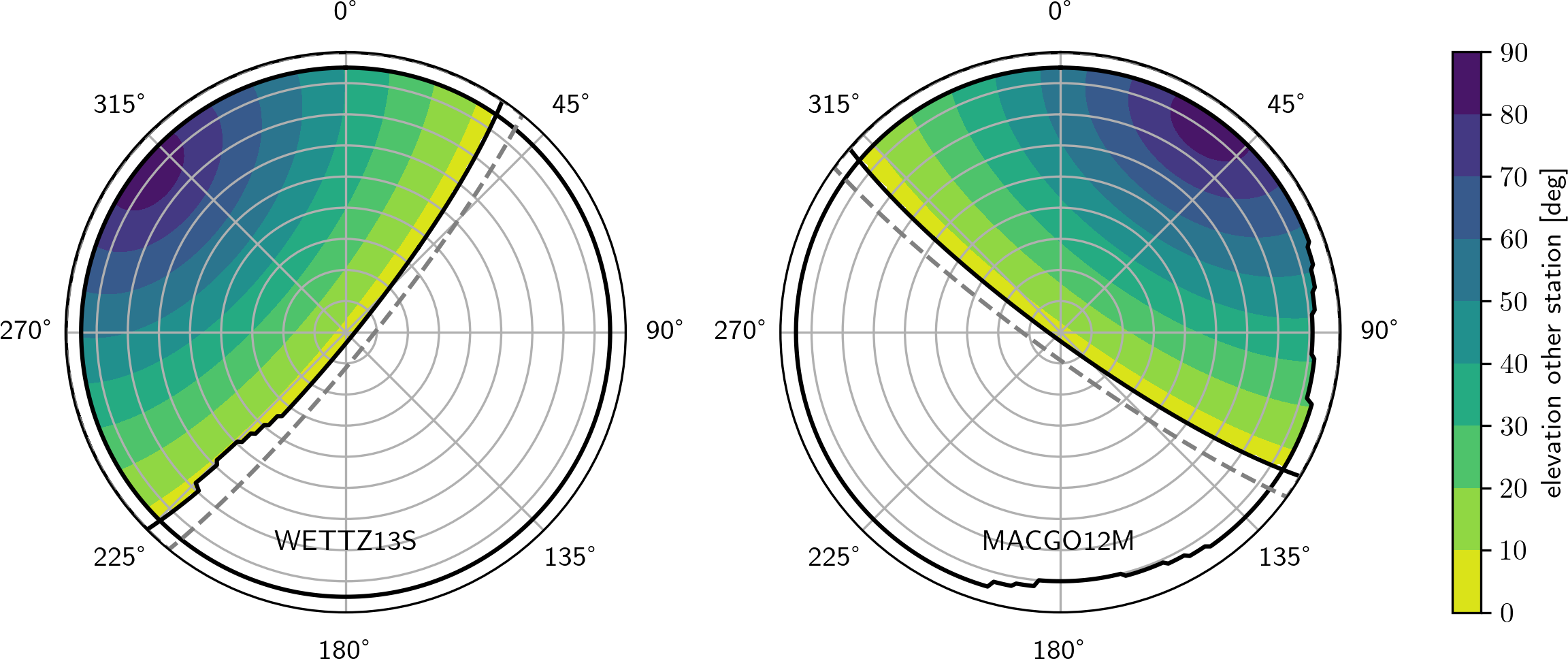}
    \caption{Mutual visibility color-coded by the elevation of the partner telescope.
    The black lines represent the station horizon masks while the dashed gray line marks the theoretical horizon.}
    \label{fig:el}
\end{figure}

Following this idea, a new VLBI observation strategy has been developed for VGOS-INT-S and applied using VieSched++ \citep{Schartner2019}.
Due to the special geometry of the Mg/Ws baseline with its baseline length of \SI{8418}{km}, observations at high elevation on one station naturally result in low elevation at the other station as depicted in Figure~\ref{fig:el}.
Although longer baselines are potentially more sensitive to UT1-UTC, they also have a limited mutually visible sky.
For example, the frequently observed \textsc{kokee/wettzell} baseline has a length of \SI{10358}{km}, resulting in a maximum observable elevation of only \SI{\sim65}{^\circ}. 
This can potentially result in a worse determination of the ZWD and thus UT1-UTC.

The new observation strategy is based on rapidly alternating between high and low-elevation scans to allow for a higher frequent ZWD determination.
Therefore, the following scan sequence is repeated:
\begin{itemize}
    \item scan with high elevation at Mg (low elevation Wz)
    \item scan without constraints
    \item scan with high elevation at Wz (low elevation Mg)
    \item scan without constraints
\end{itemize}
Thus, every other scan is especially dedicated to measuring ZWD.
The remaining scans are selected in a way to increase the sensitivity towards  UT1-UTC, e.g. by observing sources located at the corners of the mutually visible sky \citep{Schartner2021a}, or by reducing potential systematic errors caused by source-structure effects via observing a high number of different sources.

\begin{figure}
    \centering
    \includegraphics[width=\textwidth]{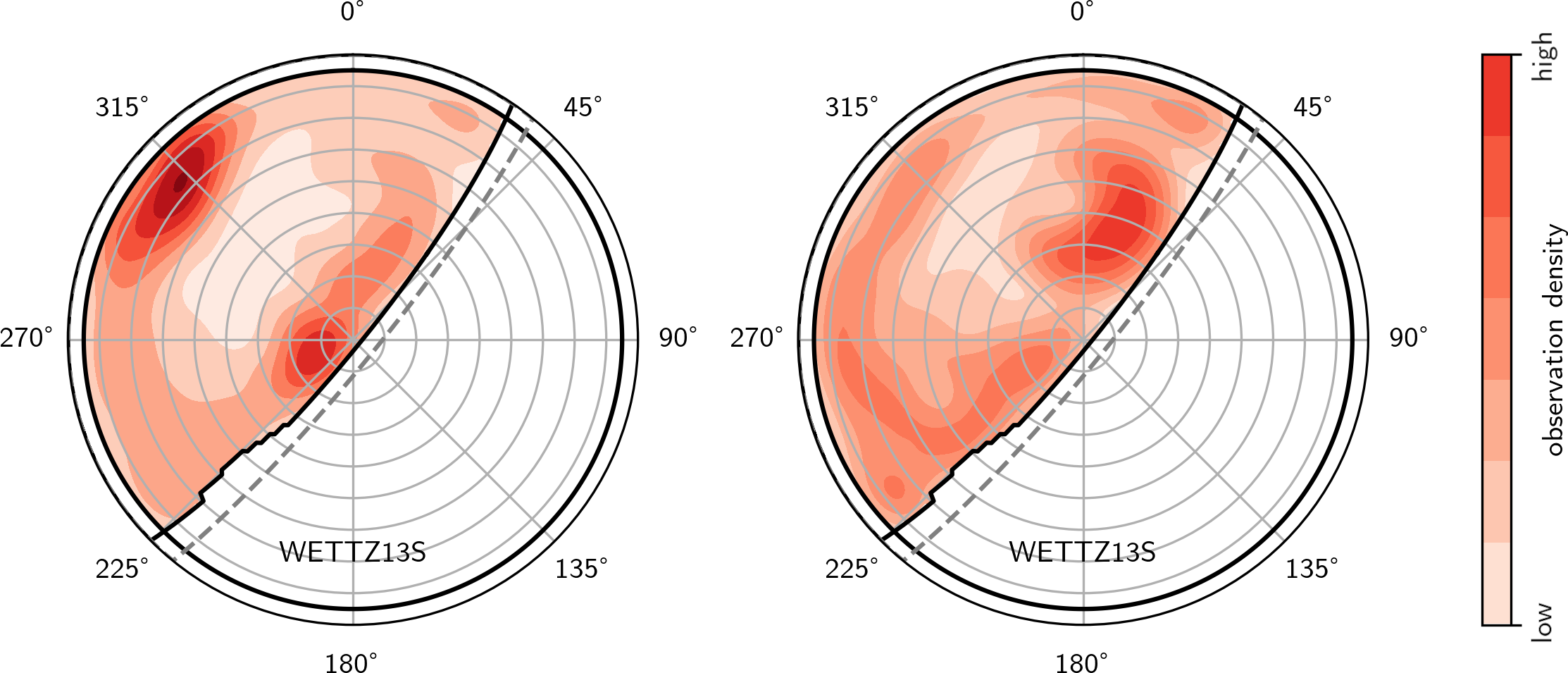}
    \caption{Scan distribution for station Ws.
    Left: new observation strategy.
    Right: old Intensive observing strategy.
    The darker the color, the higher the number of observations in this area.
    The black lines represent the station horizon mask while the dashed gray line marks the theoretical horizon.}
    \label{fig:heatmap}
\end{figure}

The effect of this special observing strategy is illustrated in Figure~\ref{fig:heatmap}, which depicts the distribution of observations in azimuth and elevation.
While the distribution is more balanced using an old observing strategy, two clear clusters are visible with the new observing strategy, one at high elevation and one at low elevation.
This confirms that the observing strategy is working as intended.

\section{Data}\label{sec:data}

The VGOS-INT-S observing program started on December 7th, 2021 with session S21341.
In the years 2021 and 2022, 25 sessions have been observed successfully.
Visibility data, geodetic databases, and results of the analysis are available at the IVS Data Centers\footnote{\url{https://ivscc.gsfc.nasa.gov/sessions/}}.

The first sessions S21341--S22011 were scheduled conservatively, using a fixed integration time of 30 seconds independent of the source brightness and thus mimicking the current 24-hour VGOS-OPS mode to gather some experience on the new baseline.
Afterward, the integration time was reduced to increase the number of observations per session, and an SNR-based observing time based on the source brightness and antenna sensitivity was utilized.
For S22018--S22053 the minimum integration time was set to 15 seconds, while it was lowered to 12 seconds in the remaining sessions while the maximum allowed integration time (except for calibrator scans) was set to 30 seconds in all sessions.
The target SNR per band was set to 15 for all sessions until S22053, while it was lowered to 10 between S22060--S22095, before being increased to 12 again for all sessions after S22109.

To troubleshoot existing hardware-related problems at the stations, two special sessions have been designed. 
First, S22277 was split into two sections.
The first 30 minutes were observed regularly with an SNR-based integration time and a minimum of 12 seconds while the second 30 minutes were scheduled using a fixed 30-second long integration time.
Second, S22284 was scheduled including \textsc{onsa13ne} (Oe; Sweden) to provide the independent baseline Mg/Oe.

Furthermore, two sessions, namely S22213 and S22215, were scheduled using a different scheduling software instead of VieSched++.

\begin{figure}
    \centering
    \includegraphics[width=1\textwidth]{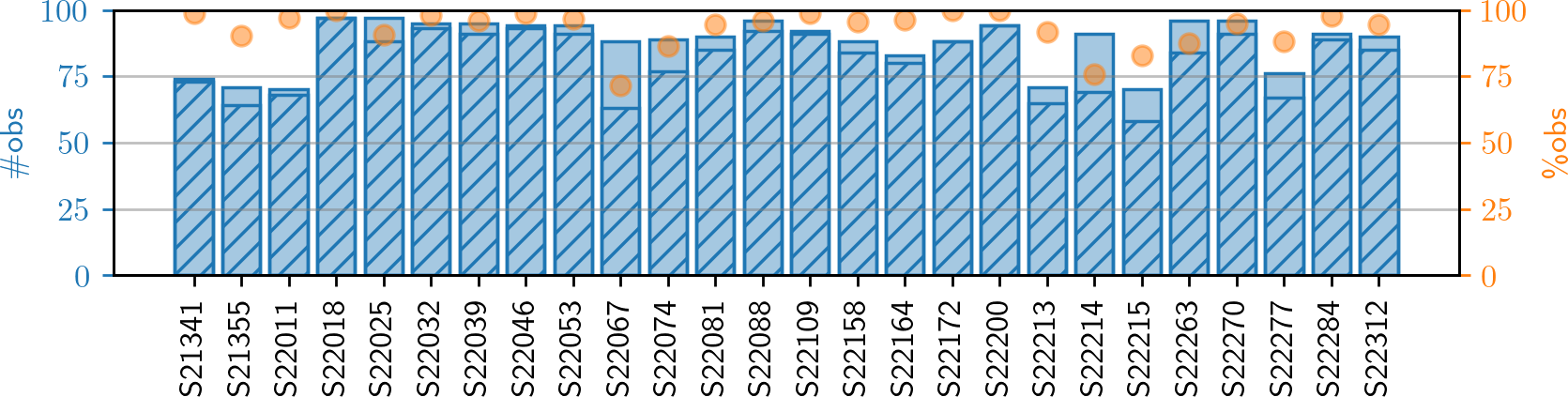}
    \caption{Number of scheduled observations (blue bars), analyzed observations (blue hatched), and percentage of analyzed observations (orange)}
    \label{fig:obs}
\end{figure}

Figure~\ref{fig:obs} depicts the number of scheduled and successful (defined as analyzed by the NASA analysis center) observations per session.
Some sessions suffered from a significant number of non-detections, mostly explained by hardware failures.
For S22067, 23 scans were not recorded at station Mg (from 19:45 until 20:00 UTC).
For S22074, 12 observations were rejected during analysis due to large residuals.
Similarly, 22 observations were rejected for S22214 and 12 observations for S22263.

A list of the most important technical problems affecting VGOS-INT-S is provided in Table~\ref{tab:issues}. 
Please note that the dates listed in this table are approximations. 
In some cases, it is no longer possible to find out exactly when a problem first occurred. 
For some of the listed problems, it is also unclear how much they affected the performance of VGOS-INT-S in general. 
\begin{table}[ht]
    \centering
    \caption{List of technical difficulties encountered during VGOS-INT-S sessions.}
    \label{tab:issues}
    \begin{tabular}{l l l l }
        \hline 
        date &  & issue & effect \\
        \hline 
        2022 Jan--May & Ws & encoder error & short downtimes \\
        2022 Feb & Mg & compressor shutdown & downtime \\
        2022 May & Mg & elevation motor coupler failed & downtime \\
        2022 May--2023 Feb & Ws & LNA failure & reduced sensitivity \\
        2022 Jun & Mg & power failure of azimuth motors & downtime \\
        2022 Jul & Mg & problems with azimuth motor stop & downtime \\
        2022 Jul--2023 Feb & Ws & 2\textsuperscript{nd} LNA failure & reduced sensitivity \\
        2022 Aug & Mg & HubPC disk failure & downtime \\
        2022 Aug--Sep & Mg & failure of M700 compressor unit & reduced sensitivity \\
        2022 Jul--Aug & Mg & phase cal signal failure & reduced sensitivity \\
        2022 Nov--2023 Feb & Ws & dewar failure & downtime \\
        2023 Jan & Mg & compressor failure & downtime \\
        \hline
    \end{tabular}
\end{table}

\section{Results}\label{sec:results}

\begin{figure}
    \centering
    \includegraphics[width=1\textwidth]{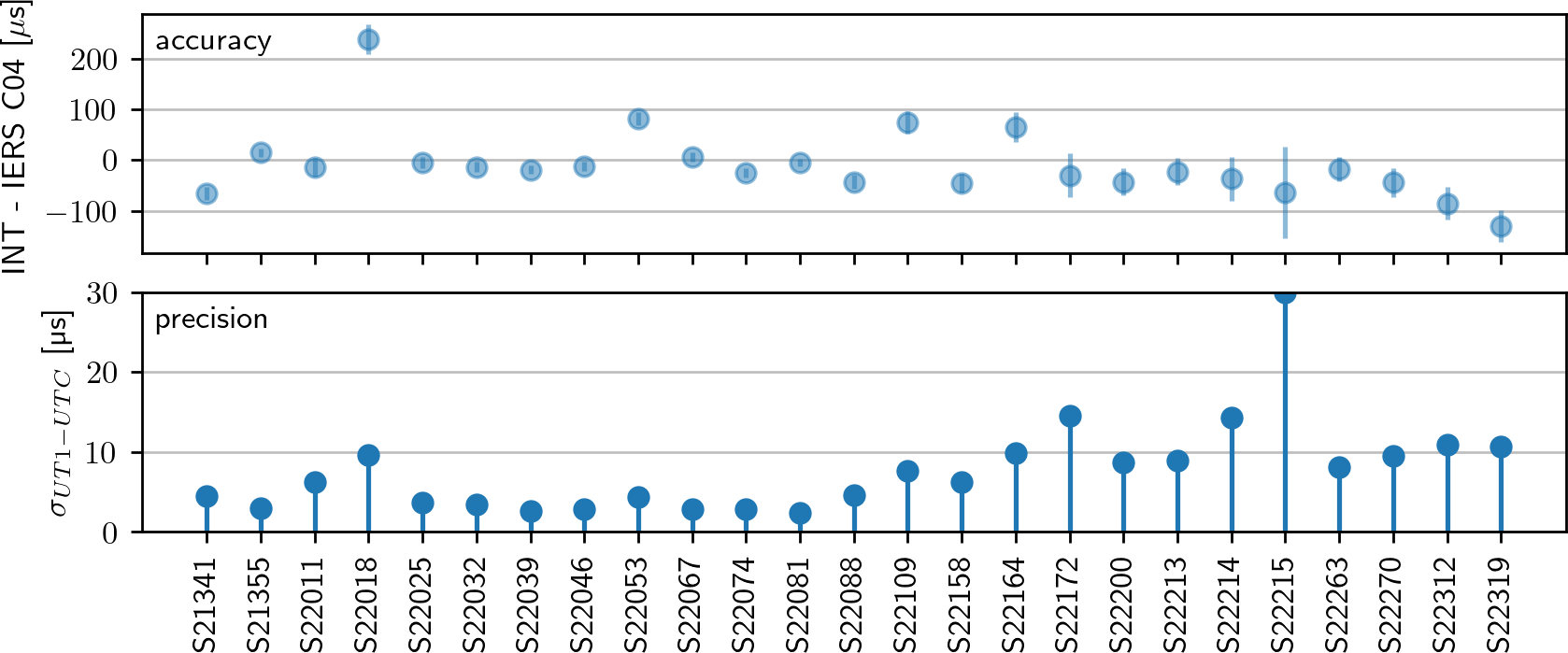}
    \caption{VGOS-INT-S accuracy (top) and precision (bottom) extracted from the GSFC analysis reports. 
    Accuracy: UT1-UTC estimate w.r.t. IERS~C04, the 3$\sigma$ value is depicted in the error bars. 
    Precision: UT1-UTC formal error $\sigma$.
    }
    \label{fig:gsfc}
\end{figure}

Figure~\ref{fig:gsfc} depicts the VGOS-INT-S precision and accuracy from the GSFC operational solutions produced with the Solve/$\nu$Solve software package.
We use two measures of UT1-UTC errors: formal uncertainties derived from the observation SNR following the law of error propagation, called here \textit{precision}, and the differences of the UT1-UTC estimates and the IERS~C04 time series, called here \textit{accuracy}.
To interpolate the daily IERC~C04 UT1-UTC values to the Intensive reference epoch, first, tidal effects with periods $<35$ days were subtracted, followed by a Lagrangian interpolation of order four and re-adding the previously subtracted tidal effects to ensure minimal interpolation error.
It is clearly visible that the formal errors between S22025 -- S22081 are significantly smaller compared to the remaining sessions.
Within this period, the average formal error $\sigma_{\text{UT1-UTC}}$ is \SI{3.1}{\micro s} while the offset w.r.t. IERS~CO4 is \SI{1.1}{\micro s} and the root mean square error (RMSE) is \SI{31.7}{\micro s}. 
For the remaining sessions, the average formal error is increased to \SI{9.8}{\micro s}, the offset w.r.t. IERS~C04 is increased to \SI{-2.1}{\micro s}, and the RMSE is increased to \SI{80.7}{\micro s}.
The increase in uncertainty can be explained by the technical problems encountered at the stations as listed in Table~\ref{tab:issues}.
Very small formal uncertainties and good agreement with IERS~C04 between S22025 and S22081 suggest that superior precision in UT1-UTC  determination at the Mg/Ws baseline can be achieved.
For comparison, during the same time, the VGOS-INT-A sessions at the \textsc{kokee12m/wettz13s} baseline achieved an average formal error of \SI{4.3}{\micro s} with an offset w.r.t. IERS~C04 of \SI{-8.6}{\micro s} and a RMSE of \SI{28.7}{\micro s}, although according to simulations, it is expected that based on the baseline geometry alone, VGOS-INT-A should be \SI{40}{\%} more sensitive towards UT1-UTC compared to VGOS-INT-S \citep{Schartner2021a}.

Here, it is to note that the GSFC operational analysis presented above does not yet make use of a more frequent estimation of ZWD enabled by the special observation strategy.
Instead, ZWD is parameterized as a constant offset only, similar to all other Intensive sessions.
This highlights that the proposed scheduling strategy, while having potential benefits through allowing a more frequent parameterization of ZWD, still provided highly accurate UT1-UTC estimates even when a traditional parameterization was used.

\begin{figure}
    \centering
    \includegraphics[width=\textwidth]{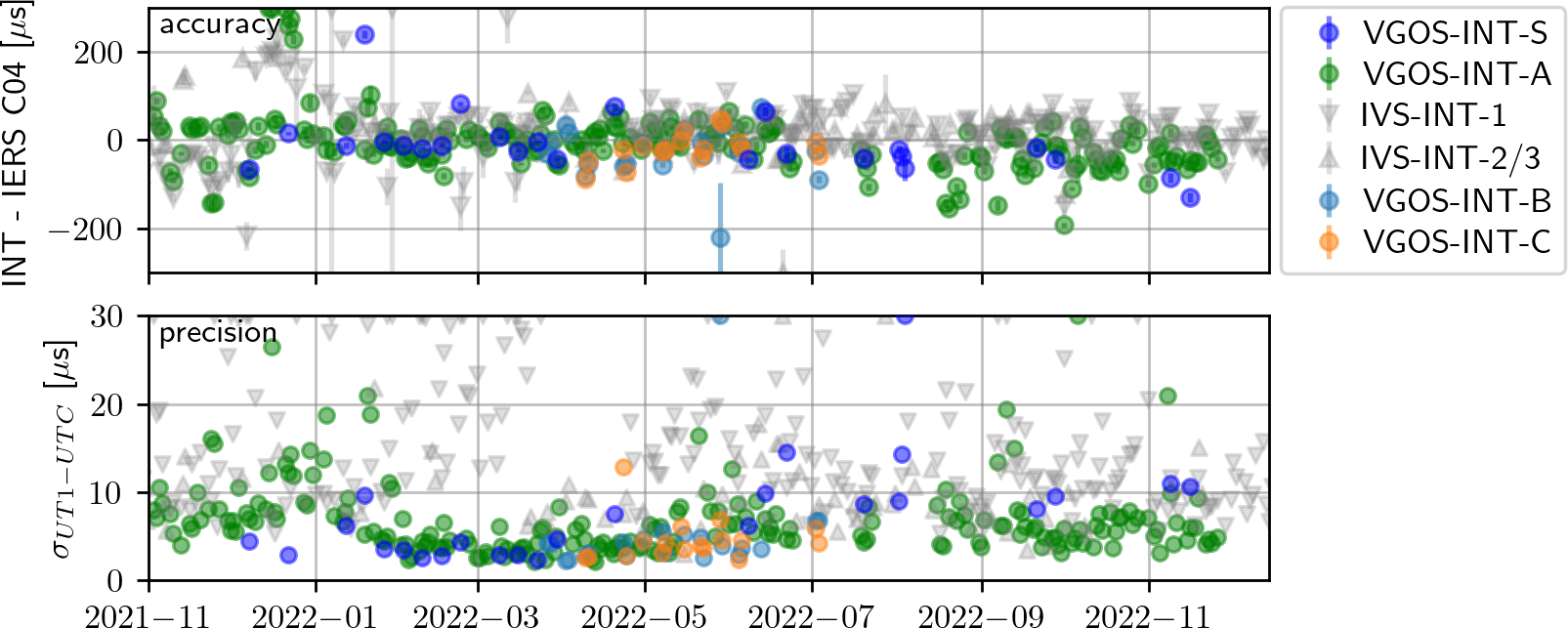}
    \caption{Comparison of various VGOS and S/X Intensive operation programs.
    Accuracy (top) and precision (bottom) were extracted from the GSFC analysis reports.
    Accuracy: UT1-UTC estimate w.r.t. IERS~C04, the 3$\sigma$ value is depicted in the error bars.
    Precision: UT1-UTC formal error $\sigma$.}
    \label{fig:int}
\end{figure}
Finally, Figure~\ref{fig:int} compares the VGOS-INT-S performance with other Intensive programs, including the previously mentioned VGOS-INT-A series, the VGOS-INT-B and VGOS-INT-C series observed between the Onsala twin stations \textsc{onsa13ne, onsa13sw} and \textsc{ishioka}, and several well established S/X series IVS-INT-1, IVS-INT-2, and IVS-INT-3 \cite{Baver2020, Schartner2022}.
From this comparison, the especially good performance of VGOS-INT-S until April 2022 is visible.
During this time, VGOS-INT-S performed best compared to all other Intensive programs.

\section{Summary}

We presented a design of a research and development VLBI observing program for the determination of UT1-UTC at a single baseline between fast slewing radiotelescopes {\sc wettz13s} (Germany) and {\sc macgo12m} (Texas, USA).

Although the telescopes suffered a number of technical failures, the results are very encouraging. 
Despite the shorter baseline length compared to more typical Intensive sessions and the resulting theoretical lower sensitivity towards UT1-UTC, the VGOS-INT-S sessions performed exceptionally well during the first part of 2022. 
We plan to continue the campaign and investigate errors of UT1-UTC determination in detail. 
We further plan to make use of a more frequent ZWD estimation during analysis, enabled by the new scheduling strategy presented in this work. 


\bibliography{sn-bibliography}

\end{document}